\input amstex
\magnification=1200
\TagsOnRight
\def\wh{\widehat}
\def\wt{\widetilde}
\def\D{\Cal{D}}

\def\ov{\overline}

\def\noi{\noindent}

\overfullrule=0pt

\def\mapab#1{\Big\downarrow}
\def\picture #1 by #2 (#3){
  \vbox to #2{
    \hrule width #1 height 0pt depth 0pt
    \vfill
    \special{picture #3} 
    }
  }

\def\scaledpicture #1 by #2 (#3 scaled #4){{
  \dimen0=#1 \dimen1=#2
  \divide\dimen0 by 1000 \multiply\dimen0 by #4
  \divide\dimen1 by 1000 \multiply\dimen1 by #4
  \picture \dimen0 by \dimen1 (#3 scaled #4)}
  }
\def\pantalon{\scaledpicture 3 in by 4.12in (pantalon scaled 500)}

\mathsurround=1pt
\tolerance=10000
\pretolerance=10000
\nopagenumbers
\headline={\ifnum\pageno=1\hfil\else\hss\tenrm -- \folio\ --\hss\fi}
\hsize=16 true cm
\vsize=23 true cm
\line{Preprint {\bf SB/F/95-234}}
\hrule
\vglue 1.5cm
\vskip 2.cm

\centerline{\bf BRST FORMULATION of 4-MONOPOLES }

\vskip 1.5cm
\centerline{  R. Gianvittorio, I.Martin and A. Restuccia}
\vskip .5cm
\centerline{\it Departamento de F\'{\i}sica, Universidad Sim\'on Bol\'{\i}var}
\centerline{\it Apartado postal 89000, Caracas 1080-A, Venezuela.}

\vskip 2.0cm

{\narrower{\flushpar {\bf Abstract.} A supersymmetric gauge invariant action 
is constructed over any 4-dimensional Riemannian manifold describing Witten's 
theory of 4-monopoles. The topological supersymmetric algebra closes off-shell.
The multiplets include the auxiliary fields and the Wess-Zumino fields in an
unusual way, arising naturally from BRST gauge fixing. A new canonical approach
over Riemann manifolds is followed, using a Morse function as an euclidean time
and taking into account the BRST boundary conditions that come from the BFV
formulation. This allows a construction of the effective action starting from 
gauge principles.\par}}

\vskip 3cm
\hrule
\bigskip
\centerline {\it e-mail: ritagian{\@}usb.ve , isbeliam{\@}usb.ve ,
arestu{\@}usb.ve }

\newpage

\noindent{\bf 1.Introduction}
\vskip .5 cm
	The non-perturbative analysis of Quantum Field Theory (QFT)is one of the
main problems that has lately been discussed in different theories. In Superstring
and Super Yang Mills theories by searching for a duality symmetry principle in 
perturbatively finite theories.In Quantum Gravity by using a loop description 
of spacetime. However, it is in Topological QFT (TQFT) that the non-perturbative analysis 
has provided the most striking results. In particular, the exact evaluation of 
correlation functions given rise to topological invariants of the base 4-manifolds, 
i.e the Donaldson invariants.
 
	This non-perturbative analysis is sucessful because of the very particular 
structure of topological quantum effective actions. One may understand this stucture 
in terms of the symmetries underlying the topological field theories. To do so, one 
may start from a lagrangian independent of the metric on the base manifold with a 
huge group of symmetries, the "topological" symmetries. Then by BRST gauge fixing one arrives 
to the off-shell BRST invariant effective action describing the topological theory.
At that level the analysis of the topological twisted supersymmetry becomes 
straightforward. Also, it is possible to see at the same time the role played by the 
"topological" symmetries in determining the linear dependence of the BRST charge on 
the conjugate momenta of the fields and ghosts. This is so in spite of the fact,that the theory 
is invariant under diffeomorphisms on the base 4-manifold. It is this feature which allows 
a complete non-perturbative analysis of the quantum theory.

	The programme of studying topological invariants by starting from a TQFT was proposed by Witten 
in [1], who found an effective twisted supersymmetric topological action for the $SU(2)$ 
instantons and obtained the Donaldson invariants from the topological observables 
of the theory.In there, he also put forward the question as to what was the gauge action 
principle associated to his effective supersymmetric action. Several gauge principles have been 
proposed in the literature [2][3], which after BRST gauge fixing lead to Witten's effective action.
The simplest one  was given in [3]. The starting point is an action independent of the metric. 
This assures that the partition function will also be independent of metrics and provides all 
the supersymmetric structure from the BRST analysis. Moreover, it allows a more general 
construction than the one proposed in [1], including the supersymmetric auxiliary fields.

	In this paper we construct in detail the gauge action principle for the topological 
theory of monopoles over 4-manifolds [4], first presented in [5]. We introduce a new canonical 
formulation over any Riemannian manifold which leads without any restrictive assumption on the 
manifold to a final covariant BRST effective action. In particular, it avoids the usual assumption 
that the base manifold is a product $R\times\Sigma_{3}$ in order to perform a canonical 
analysis. The final action we propose in section 4, is invariant under an off-shell topological 
SUSY algebra. It includes the Wess Zumino fields as well as the auxiliary fields of the 
SUSY multiplet.The Wess Zumino fields have been considered in [6] using superfields constructed 
from a general analysis of the full twisted supersymmetry algebra. However, our construction 
is simpler and allows for an explicit presentation in terms of components.In [6] also, a topological 
quantum field theory was given using the Mathai-Quillen formalism. We compare 
both results at the end of sections 3 and 4.    
\vskip .5cm
\noindent {\bf2.The gauge invariant action}
\vskip .3cm
 
	The $SU(2)$ topological quantum field theory of Witten [1] can be obtained
from a twisted version of N=2 supersymmetric Yang-Mills theory which
arises directly by BRST gauge fixing of a Lagrangian involving only the
curvature of the $SU(2)$ connection and an auxiliary 2-form [3].  As a consequence 
of the twisting there is no special requirement over the spin structure on the general 
differentiable 4-manifold and the quantum theory may be formulated starting 
from a general orientable riemannian 4-manifold. However, the construction of a gauge invariant action for
Witten monopole equations requires from the beginning the existence
of a $Spin_c$ structure over the 4-manifold, luckily this existence is assured
for any orientable riemannian manifold in four dimensions . In the case when
the second Stiefel-Whitney class of the 4-manifold is zero, i.e.
$\omega_{2}=0$, the $SO(n)$ structure group of the tangent bundle can
always be lifted to $Spin(n)$ and, hence, it is possible to define the
corresponding spin structure. In other cases when $\omega_{2}$ is reducible
modulo two of an integral cohomology class $c_1 \in H^2(X,Z)$, it is always
possible to lift $SO(n)$ to $Spin_{c}(n)= Spin(n) \times_{Z_{2}} U(1)$ and to
define a $Spin_{c}$ structure. As said before, over any orientable 4-manifold
a $Spin_{c}$ structure can always be constructed as $\omega_{2}$ is always
reducible modulo two of the integer Chern class [7]. This property is not
valid in general for manifolds of dimension $d > 4$ but holds perfectly for
orientable 4-manifolds. It is this unique property which allows the Witten
construction over a general riemannian 4-manifold. 

		The action over a general differentiable 4-manifold X we propose is given by
$$
S={1\over{8}}\int_X
(F_{\mu\nu}+B_{\mu\nu}+
{i\over{2}}\ov{M} \Gamma _{\mu\nu}
M)(F_{\rho\sigma}+B_{\rho\sigma}+
{i\over{2}}\ov{M} \Gamma _{\rho\sigma} M)dx^{\mu} \wedge
dx^{\nu} \wedge dx^{\rho} \wedge dx^{\sigma} , \tag2.1
$$
 The field  $F_{\mu\nu}$ is the curvature associated to the
$U(1)$ connection $A_{\mu}$ over a complex line bundle $L$,
$B_{\mu\nu}dx^{\mu} \wedge dx^{\nu}$ is an independent auxiliarly 2-form. $M$
and its complex conjugate $\ov{M}$ are sections of $S^+\otimes L$ and
$S^-\otimes L^{-1}$ respectively, where $L^{-1}$ is the complex conjugate
bundle of $L$ and $S^+$ is one of the irreducible parts of the spinor bundle
$S$. For any even manifold with a $Spin_c$ structure there is always a unique
spinor bundle $S$ associated to a representation of $Spin_c$ that splits into
a direct sum $S(X)= S^+(X)\oplus S^-(X)$. The Clifford
matrices $\Gamma_{\mu}$ satisfy $\{\Gamma_{\mu}, \Gamma_{\nu}\}=2
g_{\mu\nu}$.
		The  local symmetries of this action are given by the following
infinitesimal transformations:
$$
\matrix
\delta_{\lambda} A_\mu=\D_{\mu}\Lambda,&\delta_{\lambda}B_{\mu\nu}=0,&
\delta_{\lambda}M=i\Lambda M ;
\endmatrix
\tag 2.2
$$

$$
\matrix
\delta_{\epsilon}A_{\mu} = \epsilon_{\mu},&\delta_{\epsilon}M=0,
&\delta_{\epsilon}B_{\mu\nu} = -\D_{[\mu}\epsilon_{\nu{}]};
\endmatrix
\tag 2.3
$$

$$
\matrix
\delta_{\theta}M^A =\theta^A,\qquad
\delta_{\theta}B_{\mu\nu}=(-{i \over{2}}\ov{\theta}\Gamma_{\mu\nu}M
-{i\over{2}}\ov{M}\Gamma_{\mu\nu}\theta);
\endmatrix
\tag 2.4
$$
where $\Lambda$ is the local infinitesimal parameter associated to the gauge 
group $U(1)$, $\epsilon_{\mu}$
and $\theta^A$ are the local infinitesimal parameters associated to
differentiable deformations in the space of $U(1)$ connections and of
sections of $S^+\otimes L$ respectively.  We require $\epsilon_{\mu}$ to be globally
restricted by the condition that $A_{\mu}+\delta_{\epsilon}A_{\mu}$ must also be
a connection on the $U(1)$ principle bundle. $\epsilon_{\mu}$ may eliminate
any local excitation mode of $A_{\mu}$, but because of the
above global restriction it can not change the cohomology class
of $F_{\mu\nu}$. To show this point, we notice that we may always find a real parameter $\lambda$
such that $A_{\mu}+\epsilon_{\mu}$ may be written in the following way 
$$
A_{\mu}+\epsilon_{\mu} = (1-\lambda ) A_{\mu} + \lambda C_{\mu}
$$
where $C_{\mu}dx^{\mu}$ is a particular 1-form connection in $\Cal{A}$, the  convex space of
connections. We thus have
$$
\epsilon_{\mu} = \lambda (C_{\mu} - A_{\mu})
$$
this means that $\epsilon_{\mu}$ is invariant under 
transitions on the intersection of two neighborhoods on the principal
bundle. Also, under the transformation 
$$
 A_{\mu}{\to} A_{\mu}+\epsilon_{\mu}
$$
the curvature 2-form changes by
$$
F{\to}F+ d{\epsilon}
$$
but because of the invariance property of
$\epsilon_{\mu}$ under transitions on the principal bundle
$d{\epsilon}$ is an exact 2-form, i.e. $d{\epsilon}$ may change
$F$ within a cohomology class only. It can not annihilate, for example,
a Dirac monopole curvature 2-form. (2.2), (2.3) and (2.4) together with the global restriction
on $\epsilon_{\mu}$ define the gauge symmetries of our theory. That is, in the functional integral
we integrate over the gauge inequivalent classes of the fields, where two elements of the same class are
related by a finite transformation generated by the group of gauge symmetries.

	We may construct then the effective action of the gauge invariant action
(2.1) by imposing a gauge fixing condition on the 1-form $A$. However, it is difficult
to fix $A$ taking into account a global restriction on $\epsilon_{\mu}$ and at the
same time obtaining a covariant effective action under (2.2). Nevertheless,
it is possible to satisfy both conditions by imposing a gauge fixing condition on the
antisymmetric field $B_{\mu\nu}$.At first sight it would seem that we could eliminate
completely the 2-form $F$ field in the functional integral by performing a change of
integration variable on the 2-form $B$,
$$
\wt{B} = B + F
$$
ending up with a trivial theory. However, this is not the case since in the functional
integral we must integrate on the gauge equivalent classes of antisymmetric fields. We return 
to this point in eq. (3.12).

		To construct the effective action corresponding to the gauge fixing associated to
transformations (2.3), we proceed as follows. First, we notice that (2.3) contains the $U(1)$
gauge transformations (2.2). Since we would like to preserve the invariance under (2.2), we can
not fix $\epsilon_{\mu}$ completely, i.e. we must leave its longitudinal component
undetermined. Second, we need a covariant condition which may deformed to zero by
using the three remaining degrees of freedom in $\epsilon_{\mu}$. The appropiate
geometrical object is   
$$
B_{\mu\nu}^{+}\equiv{1\over{2}}(B_{\mu\nu} + {1\over{2}}g^{1/2}
\epsilon_{\mu\nu\sigma\rho}B^{\sigma\rho}) . 
$$
		We have from (2.3)
$$
\delta_{\epsilon}B_{0i}^{+} = -{1\over{2}}(\D_{0}\epsilon_{i} - D_{i}\epsilon_{0} +
{1\over{2}}g^{1/2}\epsilon_{0ijk} \D^{j}\epsilon^{k}),\tag 2.5
$$ 
given $\epsilon_{0}$ and $\delta B_{0i}^{+}$, (2.5) is a first order differential equation
in $\epsilon_{i}$ which has always a solution, allowing local infinitesimal deformations
of $B_{0i}^{+}$. Since $B_{\mu\nu}$ and $\epsilon_{\mu\nu}$ do not change under a
transition from one chart to another on the principal bundle, the deformations may be
extended globally.
		Next, since $B_{0i}^{+} = 0$ implies $B_{ij}^{+} = 0$, we may impose the covariant gauge fixing
condition
$$
B_{\mu\nu}^{+} = 0.\tag 2.6
$$
		We notice that by restricting $\epsilon_{i}$, we do not constrain the longitudinal
part of $\epsilon_{\mu}$. In fact, $\epsilon_{\mu}$ may be decomposed into
$$
\epsilon_{\mu} =\epsilon_{\mu}^{T} + \D_{\mu}\epsilon^{L}
$$
where $\D^{\mu}\epsilon_{\mu}^{T} = 0$. The longitudinal part of $\epsilon_{\mu}$ is
determined by taking
$$
\D^{\mu}\epsilon_{\mu} = \D^{\mu}\D_{\mu}\epsilon^{L}.
$$
	Given $\epsilon_{i}$, $\D^{\mu}\epsilon_{\mu}$ is still completely undetermined because
of the presence of $\epsilon_{0}$ which was not restricted by (2.6). $\epsilon^{L}$ is
the gauge parameter associated to transformations (2.2), the ordinary $U(1)$ gauge
transformations on the principal bundle.
		We may now consider the gauge fixing related to transformations (2.4). Taking into account 
the global harmonic modes over X, we may consider
$$
\align
&\Gamma^{\mu}\D_\mu M=0,\tag 2.7
\endalign
$$     
where $\Gamma^{\mu}\D_{\mu}$ is the Dirac operator that maps sections of
$S^{+}\otimes L$ to sections of $S^{-}\otimes L$.  
		The field equations associated to (2.1) are
$$
F_{\mu\nu}+B_{\mu\nu}+ {i \over{2}} \ov{M} \Gamma_{\mu\nu} M = 0 ,
$$
by using (2.6), the field equations then reduce to
$$
\align
& F_{\mu\nu}^{-}+B_{\mu\nu}^{-}=0, \tag 2.8a \\
& F_{\mu\nu}^{+}+{i\over{2}}\ov{M}\Gamma_{\mu\nu}M=0 . \tag 2.8b
\endalign
$$
From (2.6) and (2.8a) we determine the auxiliarly field $B_{\mu\nu}$. The eqs.
(2.8b) and (2.7) are the monopole equations obtained in [4].

\newpage

\noindent{\bf3.BRST gauge fixing}
\vskip .3cm
We now construct the BRST invariant action with the following
procedure  generalizing the standard one [8]. We start with a canonical formulation of (2.1) 
obtaining  a BRST invariant effective action, after a covariant gauge 
fixing and integration of the conjugate momenta in the functional integral. 
This action is manifestly covariant under general coordinate transformations.
The base manifold X we are considering is a compact
Riemannian one. Consequently, it does not satisfy the usually
assumed requirement of being globally a product $R \times \Sigma_{3}$ as is 
the case in an ordinary canonical formulation. We need to proceed
differently here.
	 We choose two points on X, namely $A$ and $B$. It is well
known that a polar Morse function always exist globally on any 4-manifold X,
with minimum and maximum height at $A$ and $B$ respectively. The
Morse function is defined by embedding X on $R_{N}$ and
considering an appropiate direction on $R_{N}$. The
projection on this direction defines a height $\tau$ as the
Euclidean "time" over X.
 
	We construct now the canonical formulation using $\tau$ as one of the local coordinates over
X. We may decompose locally X as a product $R
\times\Sigma_{3}$ between consecutive $\tau_{c}$ 3-manifolds, where $\Sigma_{3}$ is 
a compact three manifold and $\tau_{c}$ 3-manifolds define limiting manifolds that separate 
compact 3-manifolds with different number of connected components . In this way,
it is enough to decompose X into the "cilinders" determined by
the "evolution" of the disconnected parts of $\tau = constant$
submanifold $\Sigma_{3}$ between two consecutive $\tau_{c}$ manifolds , as it is 
indicated in figure 1, where $\tau_{1}$ and $\tau_{3}$ are $\tau_{c}$ 3-manifolds.
\vskip .4 cm
\centerline{\indent\pantalon
Figure 1}
\vskip .5 cm
	In the construction of a BRST invariant effective action, there
are boundary conditions that must be imposed on the BRST charge
$\Omega$ at $\tau=\tau_{i}$ and $\tau=\tau_{f}$ to have a functional
integral independent of the gauge fixing condition [8]. In our
formulation those points correspond to $A$ and $B$ respectively. The
boundary conditions are 
$$
\align
[\Omega-<\pi{\delta\Omega\over{\delta\pi}}>-<\eta_A{\delta\Omega\over{\delta\eta_A}}>-
<\ov{\eta}_A{\delta\Omega\over{\delta\ov{\eta}_A}}>-
<\mu^{1i}{\delta\Omega\over{\delta\mu^{1i}}}>-
<\mu^1{\delta\Omega\over{\delta\mu^1}}>]\mid^{\tau_{f}}_{\tau_{i}}=0,
\tag 3.1 
\endalign $$ \vskip 0.2cm
In here $<....>$ means integration on $\Sigma_{3}$. These boundary
conditions are imposed usually on the ghost fields sector associated
to the constraints that are not linear and homogeneous in the
momenta. This is the typical situation that occurs in the
Hamiltonian constraint in a diffeomorphic invariant theory. Since
(2.1) is invariant under diffeomorphisms over X, we must pay
special attention to these boundary conditions in our formulation.
The point is that we may choose $A$ and $B$ arbitrarily over
X, they are not natural limits as in a Minkowskian spacetime,
consequently, any condition on the ghost fields at these points may
ruin the possibility of a covariant construction. Fortunately, as we
will show those boundary conditions are satisfied identically
in the topological theory we propose, for any pair of points chosen.
And so, we are able to end up with a correct canonical
as well as a manifestly covariant formulation  under general
coordinate  and off-shell supersymmetric transformations.
	We discuss the canonical formulation and the boundary
conditions for other topological theories elsewhere.     
The canonical form of the action is
$$
\align
&S=\sum_{L} S_{L},
\endalign
$$
$$
\align
S_{L}=\int_{U_{L}}{}d^4x{}\;[{}&\dot{A}_i\epsilon^{ijk}(F_{jk}+B_{jk}+
{i\over{2}}\ov{M}\Gamma_{jk}M)+A_0\D_i(\epsilon^{ijk}(F_{jk}+B_{jk}+
{i\over{2}}\ov{M}\Gamma_{jk}M))+ \\
&(B_{0i} + {i \over{2}}\ov{M}\Gamma_{0i}M)\epsilon^{ijk}
(F_{jk}+B_{jk}+ {i\over{2}}\ov{M}\Gamma_{jk}M)] , \tag 3.2
\endalign
$$
The eq. (3.2) yields the
canonical conjugate momenta to $A_i$, $\pi^i$ which is a density 
under diffeomorphisms,
$$
\pi^i=\epsilon^{ijk}(F_{jk}+B_{jk}+{i
\over{2}}\ov{M}\Gamma_{ij}M).
$$

$A_0$ and $B_{0i}+ {i\over{2}}\ov{M}\Gamma_{0i}M $ are the  Lagrange
multipliers associated respectively to the constraints

$$
\align
&\phi \equiv \D_i\pi^i=0 ,\tag 3.3a\\
&\phi^i \equiv \pi^i=0 ,\tag 3.3b
\endalign
$$
the other constraints are given by
$$
\align
&\phi^A \equiv \eta_A =0,\\
&\ov{\phi}^A \equiv \ov{\eta}_A=0 ,\tag 3.3c
\endalign
$$
where $\ov{\eta}_A$ and $\eta_A$ are the conjugate momenta to $M^A$ and
$\ov{M}^A$ respectively.

All the constraints conmute, however (3.3a) and (3.3b) are not
independent. The reducibility matrix is given by
$$a\equiv{(\D_i,-1)}.\tag 3.4$$

To construct the BRST charge we follow [8] and introduce the minimal
sector of the extended phase space expanded by the conjugate pairs:

$$
(A_i,\pi^i);(M^A,\ov{\eta}_A);(\ov{M}^A,\eta_A);(C_1,\mu^1),
(C_{1i},\mu^{1i});(C_{11},\mu^{11});(C^A,\ov{\mu
}_A), \tag 3.5
$$
where we have introduced the ghost and antighost associated to the first
class constraints.

The off-shell nilpotent BRST charge is then given by:
$$
\align
&\Omega=\sum_{a}\Omega_{a}
\endalign
$$
$$
\Omega_{a}=\int_{\Sigma_{3}^{(a)}}d^3x\;
(-(\D_iC_1)\pi^i+C_{1i}\pi^i+2iC^A \ov{\eta}_A-2i\ov{C}^A \eta_A
-(\D_iC_{11})\mu^{1i}-C_{11}\mu^1) ,\tag 3.6
$$
$\Omega_{a}$ is the BRST charge for every $a$ connected component of $\Sigma_{3}$.
It is straightforward to check from (3.6) that the boundary conditions (3.1) are
satisfied identically. Hence the necessary condition to get a covariant formalism 
mentioned at the beginning of this section is fulfilled. 

We now define the non minimal sector of the
extended phase space [8]. It contains extra ghosts, antighosts and Lagrange
multipliers. First we introduce the C-fields
$$
C_m,C_{mi};\ \ \
C_{mn},C_{mni};\ \ \ \ \ \ m,n=1,2,3
$$
where at least one of the indices $m,n$ take the values 2 or  3. In
addition to these ghost, antighost and  Lagrange multiplier fields we
introduce the $\lambda$ and $\theta$ fields  (Lagrange multipliers),
also in the non minimal sector, $$
\align
&\lambda_1^0,\ \lambda_{1i}^0;\ \lambda_{1m}^0;\ \ m=1,2,3\\
&\lambda_{11}^1;\\
&\theta_1^0,\ \theta_{1i}^0;\ \theta_{1m}^0;\ \ m=1,2,3\\
&\theta_{11}^1.
\endalign
$$

In this notation the 1 subscripts denote ghost associated to a gauge
symmetry of the action, the 2 subscripts denote antighost associated to a
 gauge fixing  condition in the effective action and the 3 subscripts denote
Lagrange  multipliers associated to a gauge fixing condition.
The canonical BRST invariant effective action is then given by:
$$
\align
S_{eff}=\sum_{L}\int_{U_{L}} d^4x 
[&\pi^i\dot{A}_i+\ov{\eta}_A\dot{M}^A+\eta_A\dot{\ov{M}}^A+\mu^1\dot{C}_{1}+
\mu^{1i}\dot{C}_{1i}+\\
&\mu^{11}\dot{C}_{11}+\ov{\mu}_A\dot{C}^A+\mu_A\dot{\ov{C}}^A+
\\&\wh{\delta}(\lambda_1^0\mu^1+\lambda_{1i}^0\mu^{1i}+
\lambda_{11}^1\mu^{11}+\lambda^A\ov{\mu}_A+\ov{\lambda}^A\mu_A)+L_{GF+FP}],
\tag 3.7
\endalign
$$
where
$$
L_{GF+FP}=\wh{\delta}(C_2\chi_2+C_{2\mu\nu}\chi_2^{\mu\nu}
+C_2^{\dot{A}}\ov{\chi}_{\dot{A}}+
\ov{C}_2^{\dot{A}}\chi_{\dot{A}})+
\wh{\delta}(C_{12}\chi_{12})+
\wh{\delta}(\lambda_{12}^0\Lambda_2+
\theta_{12}^0\Theta_2),\tag 3.8
$$
is the sum of the generalizations of the Fadeev-Popov and gauge fixing
terms. In (3.8) $\chi_2$, $\chi_2^{\mu\nu}$, $\chi_{\dot{A}}$ and
$\ov{\chi}_{\dot{A}}$ are the gauge fixing
functions associated  to the constraints (3.3), while $\chi_{12}$,
$\Lambda_2$ and $\Theta_2$ are gauge fixing functions associated to the
reducibility problem. They must
fix the longitudinal part of the fields $C_{1\mu}$,$\lambda_{1}^{0}$ and
$\theta_{1}^{0}$ . Notice that $C_{2\mu\nu}$ is a self-dual density. Also, all antighosts 
in (3.8) are densities.  The BRST transformation for the canonical variables is given by 
$$
\wh{\delta}Z=(-1)^{\epsilon_z}\{ Z,\Omega \},\tag 3.9
$$
where $\epsilon_z$ is the grassmanian parity of $Z$. The BRST
transformation of the variables of the non minimal sector are given in
[8]. After integration of
the auxiliarly sector we finally choose gauge fixing functions that may be
written in a covariant form as
 $$
\align
&\chi_2=\D_{\mu}A^\mu-{\alpha\over{2}}g^{-{1\over{2}}}C_3,\\
&\chi_2^{\mu\nu}={1\over{4}}({1\over{2}}g^{-{1\over{2}}}
\epsilon^{\mu\nu\sigma\rho}B_{\sigma\rho}+B^{\mu\nu}),\\
&\chi_{12}=\D^{\mu}C_{1\mu}+{1\over{2}}
(-\ov{C}^A M_A + \ov{M}^AC_A),\\
&\ov{\chi}_{\dot{A}}=
-{i\over{2}}\D_{A\dot{A}}\ov{M}^A+\rho g^{-{1\over{2}}}\ov{C}_{3\dot{A}},\\
&\chi_{\dot{A}}=
-{i\over{2}}\D_{A\dot{A}}M^A+\rho g^{-{1\over{2}}} C_{3\dot{A}},
\tag 3.10 \endalign
$$
where $C_{1\mu}=(-\lambda_{11}^0,C_{1i})$ and $\rho$ is an arbitrary real parameter.
After elimination of all conjugate momenta in the functional integral, the
BRST transformation rules of all the remaining geometrical objects are
covariant and take the form
$$\align
&\wh{\delta}A_{\mu}=-\D_{\mu}C_1+C_{1\mu},\\
&\wh{\delta}C_1=C_{11},\\
&\wh{\delta}C_{1\mu}=\D_{\mu}C_{11},\\
&\wh{\delta}C_{11}=0,\\
&\wh{\delta}C_2=C_3,\\
&\wh{\delta}C_3=0,\\
&\wh{\delta}C_{2\mu\nu}=C_{3\mu\nu},\\
&\wh{\delta}C_{3\mu\nu}=0,\\
&\wh{\delta}C_{12}=C_{13},\\
&\wh{\delta}C_{13}=0,\\
&\wh{\delta}C^A=0,\\
&\wh{\delta}M^A=-2iC^A,\\
&\wh{\delta}C_{2}^{\dot{A}}=C_{3}^{\dot{A}},\\
&\wh{\delta}C_{3}^{\dot{A}}=0,\\
&\wh{\delta}B_{\mu\nu}^{+}= - \wh{\delta}F_{\mu\nu}^{+}- {i\over{2}}\wh{\delta}(\ov{M}\Gamma_{\mu\nu}M)
\tag 3.11
\endalign
$$
$C_{2\mu\nu}$ and $C_{3\mu\nu}$ are self dual fields.

	The BRST invariant action under the off shell nilpotent algebra (3.11) follows then from (3.7), it is given by
$$
\align
\wt{S}={1\over{8}}&\int_X
(F_{\mu\nu}+B_{\mu\nu}+
{i\over{2}}\ov{M} \Gamma _{\mu\nu}
M)(F_{\rho\sigma}+B_{\rho\sigma}+
{i\over{2}}\ov{M} \Gamma _{\rho\sigma} M)dx^{\mu} \wedge
dx^{\nu} \wedge dx^{\rho} \wedge dx^{\sigma}\\& + \int_X d^4x [\wh{\delta}(C_2\chi_2+C_{2\mu\nu}\chi_2^{\mu\nu}
+C_2^{\dot{A}}\chi_{\dot{A}}+
\ov{C}_2^{\dot{A}}\ov{\chi}_{\dot{A}})+
\wh{\delta}(C_{12}\chi_{12})], \tag 3.12
\endalign
$$
In the associated functional integral we may change variables by considering
$$
\wh{B_{\mu\nu}}= F_{\mu\nu} + B_{\mu\nu} + {i\over{2}}\ov{M}\Gamma_{\mu\nu}M
$$
ending up with the following effective action
$$
\wt{S_1}={1\over{8}}\int_X \wh{B_{\mu\nu}}\wh{B_{\rho\sigma}}dx^{\mu} \wedge
dx^{\nu} \wedge dx^{\rho} \wedge dx^{\sigma} + \int_X d^4x[\wh{\delta}(C_2\chi_2+C_{2\mu\nu}\chi_2^{\mu\nu}
+C_2^{\dot{A}}\chi_{\dot{A}}+
\ov{C}_2^{\dot{A}}\ov{\chi}_{\dot{A}})+
\wh{\delta}(C_{12}\chi_{12})],
$$
where we have to rewrite $\chi_{2\mu\nu}$ as a function of $\wh{B_{\mu\nu}}$,
$$
\chi_{2\mu\nu}= {1\over{2}}(\wh{B_{\mu\nu}^{+}} - F_{\mu\nu}^{+} - {i\over{2}}\ov{M}\Gamma_{\mu\nu}M)
$$
	We are able now to show the topological structure of the partition function $I$
associated to $\wt{S}$. In fact the only dependence on the background metric is through gauge fixing conditions.
Using the BFV theorem [8], the functional derivative of $I$ with respect to $\chi$ yields zero,
$$
{\delta{I}\over{\delta\chi}}=0
$$
where $\chi$ is any of the set of gauge fixing conditions (3.10), hence
$$
{\delta{I}\over{\delta{g_{\mu\nu}}}}= {\delta{I}\over{\delta\chi}}{\delta\chi\over{\delta{g_{\mu\nu}}}}=0
$$
showing that the theory defined by (2.1) is a topological one.

	Eliminating $\wh{B_{\mu\nu}}$ from the effective action, we obtain after functional integration
$$
\align
\wt{S_2}=\int_X d^4x g^{1\over{2}}&[-{1\over{4}}(g^{-{1\over{2}}}C_{3\mu\nu}+F_{\mu\nu}^{+} + 
{i\over{2}}\ov{M}\Gamma_{\mu\nu}M)^{2} + {1\over{4}}(F_{\mu\nu}^{+} + 
{i\over{2}}\ov{M}\Gamma_{\mu\nu}M)^{2} - g^{-{1\over{2}}}C_{2\mu\nu}\wh{\delta}\chi_2^{\mu\nu}
\\ &+g^{-{1\over{2}}}\wh{\delta}(C_2\chi_2
+C_2^{\dot{A}}\chi_{\dot{A}}+
\ov{C}_2^{\dot{A}}\ov{\chi}_{\dot{A}}+
C_{12}\chi_{12})]
\endalign
$$
$\wt{S_2}$ is invariant under the off-shell nilpotent BRST transformation (3.11).
If we eliminate $C_{3\mu\nu}$ by functional integration we obtain the action $S$,
invariant under the algebra (3.11) with the substitution
$$
g^{-{1\over{2}}} C_{3\mu\nu}= -F_{\mu\nu}^{+} - 
{i\over{2}}\ov{M}\Gamma_{\mu\nu}M ,
$$
the algebra closes now only on-shell. After some calculations involving the $\rho$
dependent terms we get the action
$$
S=S_0+S_1+S_2+S_3 , 
$$
where
$$
\align
&S_0 =\; <{1\over{2}}F^{+AB}F_{AB}^{+}+g^{\mu\nu}\D_\mu
\ov{M}^A \D_\nu M_A
+{1\over{4}}R\ov{M}^AM_A-{1\over{8}}
\ov{M}^{(A}M^{B)}\ov{M}_{(A}M_{B)}>,\tag 3.13 \\
&S_1=\; <C_2^{\mu\nu}\D_{\mu}C_{1\nu}+C_{13}\D_{\mu}C_1^{\mu}
 +C_{12}\D_{\mu}\D^\mu C_{11}>,\tag 3.14\\
\endalign
$$
$$
\align
S_2=\;
<&-C_2^{AB}(\ov{M}_{(A}C_{B)}+\ov{C}_{(A}M_{B)})
-\ov{C}_2^{\dot{A}}\D_{A\dot{A}}C^A-\ov{C}^A\D_{A\dot{A}}C_2^{\dot{A}}\\
&+{1\over{2}}(\ov{M}^AC_{1A\dot{A}}C_2^{\dot{A}}+\ov{C}_2^{\dot{A}}
C_{1A\dot{A}}M^A)
-{1\over{2}}C_{13}(\ov{C}^AM_A-\ov{M}^AC_A)\\&+2iC_{12}\ov{C}^AC_A
-{1\over{2}}\ov {M}^A\sigma_{A\dot{A}}^\mu(D_{\mu}C_{1})C_{2}^{\dot{A}}
-{1\over{2}}\ov{C}_{2}^{\dot{A}}\sigma_{A\dot{A}}^{\mu}(D_{\mu}C_1)M^A> ,
\tag 3.15\\ 
\endalign
$$
 and finally
$$
\align
S_3=\; <C_3(\D_{\mu}A^{\mu}-{\alpha\over{2}}g^{-{1\over{2}}}C_3)+C_2\D_{\mu}\D^\mu
 C_1-C_2\D_{\mu}C_1^{\mu} > \tag 3.16 \\
\endalign 
$$
where $<....>$ denotes integration on the 4-manifold X with the measure $g^{1\over{2}}$ in (3.13), and $\rho$ has 
been taken as ${1\over{8}}$ in order to cancel terms of the form $F^{+}\ov{M}M$.
In these expressions we have rewritten the objects with world indices in
terms of the corresponding ones with spinorial indices.
 
	$S_0$ corresponds to the action used by Witten in deriving the vanishing
theorems in [4]. While $S_1+S_2+S_3$ are the contributions of the ghost
and antighost fields in order to have a BRST invariant action.
The action $S_0$ agrees with the ghostless sector of the gauge fixed action
proposed in [6]. In there also, the  action is not invariant under the SUSY transformation rules
given, unless one takes 
$$
g^{-{1\over{2}}}\wh{\delta}C_{2\mu\nu}= -(F_{\mu\nu}^{+} + {i\over{2}}\ov {M}\Gamma_{\mu\nu}M)
$$
which arises from the general procedure we consider. Otherwise we agree with [6]. 
In (3.15) there are terms involving $C_{1}$ which are not present in the action given in [6]
because the latter is invariant under  BRST transformations which
close only modulo gauge transformations. While the action we
present in (3.12) is invariant under an off-shell nilpotent charge.

 In order to compare with the formulation in [1], one may perform
the change of variables:
$$
\align
&\psi_{\mu}=-iC_{1\mu},\\ &\phi=iC_{11},\\ &\eta=-C_{13},\\
&\lambda=-2iC_{12},\\ &\chi_{\mu\nu}=-C_{2\mu\nu},\\
&\mu^A=C^A, \\&v^{\dot{A}}=2iC_{2}^{\dot{A}}. \tag 3.17
\endalign $$
showing that, after the reduction from $SU(2)$ to $U(1)$, the sectors associated 
to the gauge SUSY multiplet agree. The last two changes of variables allow direct 
comparison with [6].
\vskip 1 cm
\noindent{\bf4.Topological supersymmetry}
\vskip .3cm
The action (3.12) is invariant under the off-shell BRST 
transformations (3.11) and under general coordinate transformations
over $M_{4}$ It is straightforward to obtain also a gauge invariant action.
In fact the action
$$
\wh{S}= S_{0} + S_{1} + S_{2}, \tag 4.1
$$
is still invariant under the above transformations and additionally under 
the gauge transformations on a $U(1)$ principle bundle. This is so, because
$S_{3}$ arises from the gauge fixing plus Fadeev-Popov associated to the gauge
invariance of the original action (2.1).More explicitly,
$$
S_{3}= <\wh{\delta}(C_{2}\chi_{2})>, \tag 4.2
$$
and hence it is BRST invariant on its own as well as under general coordinate 
transformations.

	In [6], the full twisted supersymmetric algebra was constructed using 
a redefinition of the $SO_4$ generators in which an identification of the isospin indices
as right handed spin indices was performed
$$
\wt{J_{AB}}= J_{AB} -2i T_{AB}, \tag 4.3
$$
It was shown there that there is a unique (up to a global factor) linear combination
of the SUSY generators $Q_{a\alpha}$ such that it behaves as a scalar, under the new 
$SO_4$ generators (4.3), satisfying $Q^{2}=0$.
	(3.11) is a realization of this nilpotent supersymmetric generator with an unusual 
Wess-Zumino auxiliary field structure with respect to the standard SUSY one [9].

	In [6], an off-shell superfield realization of twisted supersymmetry resembling the Euclidean 
SUSY realization was introduced in terms of odd grassmannian coordinates $\theta$ and 
$\theta^{\alpha\beta}$. However, to obtain the component action a Wess-Zumino gauge is 
introduced.The SUSY algebra may then close only modulo field dependent gauge transformations.
The Wess-Zumino auxiliary fields introduced in the superfields $W$ and $V$ are the twisted 
version of the usual ones [9]. What we show in (3.11) is that the off-shell closure of the 
subalgebra generated by the nilpotent charge $Q$  may be obtained by introducing only the $C_{1}$ auxiliary 
field. Moreover (3.12) is the invariant SUSY action associated to that realization.
  
	We show now how to obtain the twisted SUSY algebra, in a Wess-Zumino gauge, from our nilpotent BRST
algebra. Let us consider the transformation law for $A_{\mu}$. We define
the SUSY transformation by
$$
\delta A_{\mu} :=\wh{\delta}A_{\mu}\;\mid_{C_{1}=0} ,\tag 4.4
$$
we then have from (3.10)
$$
\align
\delta A_{\mu}&=C_{1\mu},\\
\delta \delta A_{\mu}&=\delta C_{1\mu}=\D_{\mu}C_{11}.\\ \tag 4.5
\endalign
$$

The SUSY algebra thus closes up to a gauge transformation generated by
$C_{11}$ as required.

	The SUSY transformation for $M^A$ may be obtained by considering an
equivalent BRST formulation to (3.5). Instead of considering the
constraint (3.2a) we may take equivalently
$$
\D_{i}\pi^i+M^{A}\ov{\eta}_{A}+\ov{M}^{A}\eta_{A}=0, \tag 4.6
$$
The associated BRST charge is then given by
$$
\align
\Omega_{a}=\int_{\Sigma_{3}^{(a)}}d^3x\;
(&-(\D_iC_1)\pi^i+C_{1i}\pi^i+2iC^A \ov{\eta}_A-2i\ov{C}^A \eta_A
-(\D_iC_{11})\mu^{1i}-C_{11}\mu^1 \\&+C_{1}M^A \ov{\eta}_A+C_{1}\ov{M}^A
\eta_A+{i\over{2}}C_{11}M^A \ov{\mu}_A \\&-{i\over{2}}C_{11}\ov{M}^A
\mu_A-C^{A}\ov{\mu}_{A}C_{1}-\ov{C}^{A}\mu_{A}C_{1})\\ \tag 4.7
\endalign
$$
comparing with (3.5) we see some other terms coming from the new choice of
constraints. The BRST charge is again nilpotent when acting on the
configuration space of the fields after the elimination of the auxiliary
ones. The nilpotent BRST transformation laws are now
$$
\align
\wh{\delta}M^A&=-2iC^A-C_{1}M^{A},\\
\wh{\delta}C^A&= {i\over{2}}C_{11}M^{A}-C^{A}C_{1},
\endalign
$$
there are analogous changes for $\ov{M}^A$ and $\ov{C}^A$, while the
transformation law for the other fields are as in (3.11).
We define as before the SUSY transformations of $M^A$ and $C^A$. We have
$$
\align
\delta M^{A} &:=\wh{\delta}M^{A}\;\mid_{C_{1}=0}\\
\delta C^{A} &:=\wh{\delta}C^{A}\;\mid_{C_{1}=0}
\endalign
$$
We then obtain
$$
\align
\delta\delta M^{A} &=-2i\delta C^{A}= C_{11}M^{A}\\
\delta\delta C^{A} &={i\over{2}}C_{11}\delta M^{A}= C_{11}C^{A}
\endalign
$$
as required. As shown the full SUSY algebra results from our nilpotent BRST charge.
The BRST formulation using the combination of constraints (4.6) we have considered, 
leads to an effective action which may be obtained from (3.12) by a canonical
change of coordinates in the original extended symplectic geometry. It is then a completely 
equivalent formulation. 
\vskip .3cm
\noindent{\bf5.Conclusions}
\vskip .3cm

		In summary, we introduced a topological action with a large class of
local symmetries, whose field equations are the Witten monopole
equations found in [4]. By following a covariant gauge fixing procedure
we obtained a covariant BRST invariant effective action . The BRST
generator obtained is nilpotent off-shell. The canonical construction of
the nilpotent BRST charge has been carried out without any further
requirements on the base manifold beyond those assumed for the set up of
action (2.1). This construction uses particular properties of the
BRST charge for this topological theory, allowing a general canonical 
analysis on any compact 4-manifold $X$ without the standard assumption 
that $X=R\times \Sigma_{3}$. Finally we show how the twisted  N=2 supersymmetric algebra used to get the
Witten topological theory may be directly obtained from our
nilpotent BRST charge.

	 One of the main consequences of the existence of
action (2.1) would be the possibility of relating the $SU(2)$ topological
quantum field theory [1] directly to  Witten's 4-monopoles theory [4]. In
fact, the action (2.1) could be obtained by a partial gauge fixing that
breaks the $SU(2)$ invariance to a $U(1)$ in the action already obtained in
[3] for the $SU(2)$ topological theory together with some extra assumptions 
that we will discuss in a forthcoming communication.  This would allow to compare
directly the correlation functions of both topological theories by using
the BFV theorem. This procedure seems interesting since does not use
explicitly the duality relation between both theories found in [4] over a flat 
background nor it does use any supersymmetric argument.

\newpage
\parindent=0pt

\centerline{\bf APPENDIX}
\vskip 1cm
Here we summarize the conventions used in this paper.
To raise and lower spinor indices we use the matrix $C_{AB}$,
$$
C_{AB} = (\tau_2)_{AB}
$$
where $\tau_i$, $i=1,2,3$ are the Pauli matrices:
$$
\tau_{1} = \pmatrix 0 \;\;\;\;\; 1 \\
1 \;\;\;\;\; 0 \\
\endpmatrix, \;\; \tau_{2} = \pmatrix 0 \;\; -i \\
i \;\;\;\;\; 0 \\
\endpmatrix, \;\; \tau_{3} = \pmatrix 1 \;\;\;\;\; 0 \\
0 \;\; -1 
\endpmatrix 
$$
The inverse of $C_{AB}$ is given by $C^{AB}$ so that
$$
C^{AB}C_{DB}=\delta_{D}^{A}
$$
Lowering and raising spinors $\psi$ follow the rules
$$
\align  
&\psi_A = \psi^B C_{BA}\\
&\psi^A  =  C^{AB}\psi_B
\endalign
$$
To have a real $S_{eff}$ and a hermitian BRST charge $\Omega$ we define
$$
\align 
&\ov{M}^A = M_A^*\\
&\ov{M}_A = -M^{A *}\\
&\eta^A = -\ov{\eta}_A^*\\
&\eta_A = \ov{\eta}^{A*}\\
\endalign
$$
and for the even field $C^A$ 
$$
\align 
&\ov{C}_A = C^{A*} \\
&\ov{C}^A = -C_{A}^{*}
\endalign 
$$
For even and odd objects with spinorial indices, we have the rule:
$$
\psi_A \ov{\phi}^A = - \psi^A \ov{\phi}_A
$$
We also use the set of four matrices $\sigma_{\mu}$, $\mu$= 0,1,2,3, as
$$
\sigma_{0}= \pmatrix 1 \;\;\;\;\; 0 \\
0 \;\;\;\;\; 1 \\
\endpmatrix, \;\; \sigma_{a}= -i\tau_{a}^T, \; a=1,2,3 .
$$
and the matrices 
$$
\wt{\sigma}_{\mu}^{A\dot A}= C^{AB}C^{\dot A\dot B}\sigma_{\mu B\dot B}= -C^{AB}\sigma_{\mu B\dot B}C^{\dot B\dot A},
$$
$$
\wt{\sigma}_{0}= -\pmatrix 1 \;\;\;\;\;0 \\
0 \;\;\;\;\; 1 \\
\endpmatrix, \;\; \wt{\sigma_{a}}= -i\tau_{a}, a=1,2,3.
$$
then we define ${\sigma^{\mu\nu}}_A ^B$ by:
$$
2{\sigma^{\mu\nu}}_A^B = g^{\mu\nu} \delta_A^B  + \sigma^\mu _{A\dot A}\wt{\sigma}^{\nu B\dot A}
$$
they satisfy the following properties:
$$
\align
i&)\qquad \qquad{\sigma^{\mu\nu}}_A^B \sigma^\lambda _{B\dot B} = {1\over{2}} g^{\lambda[\mu} \sigma^{\nu]}_{A\dot B} + {1\over{2}}\epsilon^{\mu\nu\lambda\rho}\sigma_{\rho \,A\dot B}\\ 
ii&)\qquad\qquad{\sigma^{\mu\nu}}_A^B{\sigma^{\lambda\rho}}_B^A = -{1\over{2}}g^{\lambda[\mu} g^{\rho\nu]} - {1\over{2}}\epsilon^{\mu\nu\lambda\rho}  \\
iii&)\qquad\qquad{\sigma^{\mu\nu}}_{AB}{\sigma_{\mu\nu}}^{CD} = \delta_{(A}^C \delta_{B)}^D 
\endalign
$$
where $\epsilon^{0123} = -1$ and $A^{[\mu} B^{\nu]} = A^{\mu} B^{\nu} - A^{\nu} B^{\mu}$. 
The 2-form $B$ may be decomposed into its self-dual and anti-self-dual parts:
$$
B_{\mu\nu} = B_{\mu\nu}^+ + B_{\mu\nu}^-
$$
where
$$
B_{\mu\nu}^\pm = {1\over{2}}(B_{\mu\nu} \pm {1\over{2}}\epsilon_{\mu\nu\rho\sigma}B^{\rho\sigma})
$$
we then have
$$
\align
&B_{AB} = a\,\sigma_ {\rho\lambda AB} B^{\rho\lambda}\\
&B_{\mu\nu}^+   = b\,\sigma_ {\mu\nu}^{AB} B_{AB}
\endalign
$$
where 
$2 ab=1$, so we may choose $a ={1\over{2}}$ and $b=1$.

The full covariant derivatives in terms of the $U(1)$ connection 
and the covariant derivatives on the manifold $X$ are given by:
$$
\D_{A\dot A} = \sigma^\mu_{A\dot A} \D_\mu
$$
where
$$
\D_\mu^\pm = \nabla_\mu   \pm iA_\mu
$$
and $\nabla_\mu$ is the covariant derivative on the manifold $X$.

\newpage

\vskip 1cm
\noi
{\bf REFERENCES}
\vskip .3cm

\item{[1]}E. Witten, {\it Commun. Math. Phys.} {\bf 117}
(1988) 353.
\item{[2]}J.M.F. Labastida and M. Pernici, {\it Phys. Lett.} {\bf B212}
(1988) 56.
\item{}L. Baulieu and I. M. Singer, {\it Nucl. Phys.} (Proc.
Suppl.) {\bf 5B} (1988) 12.
\item{}Y. Igarashi, H. Imai, S. Kitakado and H. So, {\it Phys. Lett.}
{\bf B227} (1989) 239.
\item{}C. Arag\~{a}o and L. Baulieu, {\it Phys. Lett.}
{\bf B275} (1992) 315.
\item{[3]}R. Gianvittorio, A. Restuccia and J. Stephany, {\it Phys.
Lett.} {\bf B347} (1995) 279.
\item{[4]}E. Witten,{\it Math. Research Lett.} {\bf 1} (1994) 769.
\item{}N. Seiberg and E. Witten,{\it Nucl. Phys.} {\bf B426} (1994) 19.
\item{}S. Hyun, J. Park, J-S. Park, hep-th/9508162. 
\item{[5]}R. Gianvittorio,I. Martin and A. Restuccia, talk given at LASSF II, 
USB, Caracas October 1995   
\item{[6]}M. Alvarez and J.M.F. Labastida, {\it Nucl. Phys.} {\bf B437} (1995) 356.
\item{}J.M.F. Labastida and Mari\~no, SLAC-PUB-US-FT 3/95.
\item{}J.M.F. Labastida and Mari\~no, hep-th/9507140.
\item{[7]}F. Hirzebruch and H. Hopf, {\it Math. Ann.} {\bf 136} (1958) 156.
\item{[8]} M.I.Caicedo and A. Restuccia, {\it Class. Quan. Grav} {\bf 10 } (1993) 833.
\item{}M.I.Caicedo and A. Restuccia, {\it Phys. Lett.} {\bf B307 } (1993) 77.
\item{[9]} M.Sohnius, {\it Phys. Rep.}{\bf 128} (1985) 39-204. 
\end